\documentclass[preprint,aps]{revtex4}

\usepackage{graphicx}
\usepackage{dcolumn}
\usepackage{bm}

\begin{document}

\preprint{}

\title{Berry Phase in Neutrino Oscillations }

\author{$^{1,2}$Xiao-Gang He\footnote{hexg@phys.ntu.edu.tw},
$^1$Xue-Qian Li\footnote{lixq@nankai.edu.cn} $^{3}$Bruce H. J.
McKellar\footnote{mckellar@physics.unimelb.edu.au}, $^4$Yue
Zhang\footnote{yzhang@pku.edu.cn }}
\affiliation{%
$^1$ Department of Physics, Nankai University, Tianjin\\
$^2$NCTS/TPE, Department of Physics, National Taiwan University,
Taipei\\
$^3$School of Physics, University of Melbourne, Parkville,
Victoria\\
$^4$Department of Physics, Peking University, Beijing}

\date{\today}

\begin{abstract}
We study the Berry phase in neutrino oscillations for both Dirac
and Majorana neutrinos. In order to have a Berry phase, the
neutrino oscillations must occur in a varying medium, the
neutrino-background interactions must depend on at least two
independent densities, and also there must be CP violation if the
neutrino interactions with matter are mediated only by the
standard model W and Z boson exchanges which implies that there
must be at least three generations of neutrinos. The CP violating
Majorana phases do not play a role in generating a Berry phase. We
show that a natural way to satisfy the conditions for the
generation of a Berry phase is to have sterile neutrinos with
active-sterile neutrino mixing, in which case at least two active
and one sterile neutrinos are required. If there are additional
new CP violating flavor changing interactions, it is also possible
to have a non-zero Berry phase with just two generations.
\end{abstract}

\maketitle

\section{ Introduction}

In the past decade, great progress has been made in the study
of neutrinos. One of the most important achievements is the
observation of neutrino oscillations\cite{oscillations}. If
different species of neutrinos have different masses and also mix
with each other neutrino oscillations can occur in
vacuum\cite{pmns} and also in matter\cite{msw}. The
oscillation effects are closely related to the phases of neutrino
fields.

In quantum mechanics, particle wave functions satisfy the Schr\"odinger
equation. Once the Hamiltonian of the system is known, the
evolution of a quantum system is determined.
Berry has shown that if the Hamiltonian depends
on time via a set of adiabatic parameters, besides the usual
dynamic phase, a non-dynamic phase (the Berry phase) will also be
developed\cite{berry}. Neutrino oscillations in matter,
where the varying matter density plays the role of
the adiabatic parameters, fit such
a situation nicely. When the neutrinos move through a
medium, a Berry phase is then expected to be generated.

In this work we study the conditions with which a Berry phase can
be developed in neutrino oscillations when passing through a
medium. We will consider Dirac and Majorana neutrinos with both
active and active-sterile neutrino mixing, and also study cases
with new CP violating flavor changing interactions. We find that
in order to generate a Berry phase ,
\begin{itemize}
\item  there must be $CP$ violation,

\item but  the CP violating phases peculiar to Majorana neutrinos
do not contribute,

\item there must be three neutrino generations if neutrinos
interact with the medium only through W and Z exchanges, and

\item the neutrino-background matter interaction must depend on
two independent densities, such as the electron and neutron
densities (which requires interactions beyond the standard model,
or the introduction of sterile neutrinos), or the electron and
background neutrino densities (which is possible in such
astronomical objects as supernovae).
\end{itemize}

We find that active and sterile neutrino mixing can provide a
natural realization of all the conditions listed above due to
different interactions of the Z boson with active and sterile
neutrinos. If there are new interactions of neutrinos with matter,
it is possible to have a Berry phase with just two neutrinos.

Several authors have considered the phase properties in neutrino
oscillations\cite{naumov,wrong}. Our discussions follow Berry's
definition of Berry phase in terms of a parametric dependence of
the Hamiltonian, and the phase depends on motion through a loop in
parameter space\cite{berry}. This is in contrast to
Ref.\cite{wrong} who used a different definition due to
Ref.\cite{anandan} which will not be discussed here.

The conditions of generating a Berry phase with Dirac neutrino
oscillation in matter with the standard W and Z interactions have
been discussed in Ref.\cite{naumov}. Our results generalize the
discussions in Ref.\cite{naumov} to include Majorana neutrinos,
active and sterile neutrino mixing, and new interactions. Specific
examples which can realize all the conditions are discussed.

\section{Dirac Neutrino Oscillation and Berry Phase}

We start with the study of the Berry phase for Dirac neutrinos
going step by step to identify the origin of Berry phase. The
discussions in this section also serve to set up notations for the
analysis in the other sections.
 Neutrino
oscillations are due to the mismatch of mass and weak interaction
eigenstates of the interaction Lagrangian. For Dirac neutrinos the
relevant Lagrangian, in the weak eigenstate basis where the
charged leptons are already in their mass eigenstate basis, is
given by
\begin{eqnarray}
L &=& \bar \nu_L i \gamma^\mu \partial_\mu \nu_L + \bar \nu_R i
\gamma^\mu \partial_\mu \nu_R -\bar \nu_R M \nu_L -\bar \nu_L
M^\dagger \nu_R\nonumber\\
&-&{g\over 2 \cos\theta_W } \bar f \gamma_\mu(g^f_V -
g^f_A\gamma_5) f Z^\mu
 - {g\over \sqrt{2}} \bar l_L
\gamma^\mu \nu_L W_\mu^- + h.c., \label{ldirac}
\end{eqnarray}
where $g_V = T^f_3 - 2 \sin^2\theta_W Q_f$ and $g_A = T^f_3$ with
$f$, $T^f_3$ and $Q_f$  the Standard Model left-handed fermions
and their corresponding isospin and electric charges.

The mis-match of mass and weak eigenstates means that the neutrino
mass matrix $M$ is a non-diagonal matrix. $M$ can be diagonalized
to a diagonal mass matrix $\hat M$ by a bi-unitary transformation,
so that $M = U^\prime \hat M U^\dagger$. Here $U^\prime$ and $U$
are unitary matrices. When writing the Lagrangian in the mass
eigenstate basis $\nu^m_L = U^\dagger \nu_L$ and $\nu^m_R =
U^{'\dagger} \nu_R$, the form of the first four terms does not
change, with $M$ replaced  by $\hat M$, but the W interaction term
in the above equation changes to $-(g/\sqrt{2}) \bar l^m_L
\gamma^\mu U \nu^m_L$. $U$ is the PMNS\cite{pmns} mixing matrix,
and $U^{'}$ is the equivalent matrix for the right handed
neutrinos which plays no further role in our analysis as it does
not enter the standard model interactions. By redefining the
lepton field phases, for n generations of leptons, the mixing
matrix $U$ can be parameterized by $n(n-1)/2$ rotation angles and
$(n-1)(n-2)/2$ CP violating phases.

A weak eigenstate of neutrino $\nu_i$ of energy $E$ produced in
association with a charged lepton would propagate in vacuum as

\begin{eqnarray}
\nu_i(t) = \sum_j U_{ij}e^{-ix\cdot p_j}\nu^m_j(0) =\sum_{jk}
U_{ij}U^*_{kj}\nu_k(0)e^{-ix\cdot p_j}. \label{vacuum}
\end{eqnarray}
Here $p_j = (E, \vec p_j)$ and $|\vec p| = \sqrt{E^2-m^2_j}
\approx E - m^2_j/2E$.

The probability amplitude for observing a $\nu_k$ neutrino at time
$t$ after the creation of a $\nu_i$ neutrino,
a distance $L$ away in the direction of propagation, is given by

\begin{eqnarray}
A(\nu_i \to \nu_k) = \sum U_{ij}U^*_{kj}e^{-ix\cdot p_j}\approx
e^{-iE(t-L)}\sum U_{ij}U^*_{kj} e^{-im^2_j L /2E}.
\end{eqnarray}
$|A(\nu_i \to \nu_k)|^2$ is the transition probability for $\nu_i$
to $\nu_k$.

The phases $-x\cdot p_j $ in the above are usually referred as
dynamic phases. Since we follow Berry's definition of Berry phase
in terms of a parametric dependence of the Hamiltonian, and the
phase depends on motion through a loop in parameter
space\cite{berry}, the dynamic phase $-x\cdot p$ will not
contribute to the Berry phase. In order to study the Berry phase
defined in Ref.\cite{berry} one must separate the effects of the
dynamic phases and identify phases which depend on some slowly
varying adiabatic parameters in a given system. In the case of
neutrino oscillation, we find that the slowly varying matter
background can be identified as the adiabatic parameter.

In medium with a finite matter density neutrino propagation is
different to that in vacuum because of the interaction of
neutrinos with background matter. One can obtain the equation of
motion in this case by integrating out the W and Z in eq.
(\ref{ldirac}) obtaining the four fermion interaction term $-
\sqrt{2} G_F \bar \nu_L \gamma^\mu (j_\mu N + j'_\mu N')\nu_L$ in
the Lagrangian. In the Standard Model $N$ and $ N'$ are diagonal
matrices representing interaction of neutrinos with the medium due
to $W$ and $Z$ exchange respectively. For the purpose of
discussing neutrino oscillations in the sun and in the earth which
are unpolarized, $j_\mu = \bar e \gamma_\mu e$ is the electron
number density current and $N = \mbox{diag}(1,0,0)$. This is due
to $W$ exchange between the neutrinos and the background
electrons. $N'$ is a unit matrix with $j'_\mu = \sum_f \bar
f\gamma_\mu f (T^f_3 - 2 Q_f \sin^2\theta_W)$ generated by Z
exchange.  The term proportional to $N'$ does not affect the
mixing of the active neutrinos, and will therefore be ignored
until section \ref{sterile}. A non-trivial effect will show up
there when we discuss oscillations with sterile neutrinos.

The equations of motion are given by
\begin{eqnarray}
&&i\gamma^\mu \partial_\mu \nu_L - M^\dagger \nu_R - \sqrt{2}G_F
j_\mu \gamma^\mu N \nu_L =0, \nonumber\\
&& i\gamma^\mu \partial_\mu \nu_R - M\nu_L=0.
\end{eqnarray}

For a static case, $j_\mu = \rho(x) g_{\mu 0}$. Writing $\nu_L =
e^{-iEt}\psi(x)$ for fixed energy $E$ much larger than $m_i$, and
assuming slowly varying density $(\rho E)^{-1} d\rho/dx << 1$, the
equation of motion for $\psi$ for neutrino beam travelling along x
direction is given by\cite{hep}
\begin{eqnarray}
[E^2 - M^\dagger M + \partial_x\partial_x - 2\sqrt{2} G_F \rho
E
N]\psi(x)=0.
\end{eqnarray}
We have used the approximation
$\gamma^0 \gamma^i(-i\partial_i) \psi(x) \approx -E \psi(x)$ for
$E >> m_i$.

The above equation can be written as $(i{d/dx} + (E^2 - M^\dagger
M  - 2\sqrt{2} G_F \rho E N)^{1/2}) (i{d/ dx} - (E^2 - M^\dagger M
- 2\sqrt{2} G_F \rho E N)^{1/2})\psi(x) =0$, where the assumption
of slow variation is again used. For not too large density
$\rho(x)$, $(E^2 - M^\dagger M - 2\sqrt{2} G_F \rho E N)^{1/2}
\approx E-M^\dagger M/2E - \sqrt{2} G_F \rho N$, and one obtains
the usual equation of motion for neutrinos in matter\cite{msw},

\begin{eqnarray}
i{ d\over dx} \psi(x) = -H \psi(x),\;\;\;\;H = E-{M^\dagger M\over
2 E} - AN, \label{medium}
\end{eqnarray}
where $A = \sqrt{2} G_F \rho$.

One can write $H$ in the following form
\begin{eqnarray}
H &=& \tilde E - \tilde A,\nonumber\\
\tilde E &=& E - {1\over 2 E} {m^2_1+m^2_2 + m^2_3\over 3} -
{A\over 3},\;\;
\tilde A = {1\over 2 E}\left (
\begin{array}{lll} a_{11}&a_{12}&a_{13}\\
a_{12}^*& a_{22}&a_{23}\\
a_{13}^*&a_{23}^*& a_{33}\\
\end{array}\right ) \nonumber\\
a_{ij} &=& {1\over 3}(\Delta m^2_{21} - \Delta
m^2_{32})\delta_{ij} - \Delta m^2_{21} U_{i1}U^*_{j1} +\Delta
m^2_{32} U_{i3}U^*_{j3}\nonumber\\
 &+&{1\over 3} 2E A
(2\delta_{i1}\delta_{j1}
-\delta_{i2}\delta_{j2}-\delta_{i3}\delta_{j3}).
 \label{matrix}
\end{eqnarray}
Note that $\tilde A$ is a traceless Hermitian matrix.

Writing $ \psi = e^{i\int^x_0 \tilde E dx} \tilde \psi$, we find
that $\tilde \psi$ satisfies the following equation of motion
\begin{eqnarray}
i {d\over dx} \tilde \psi = \tilde A \tilde \psi.
\end{eqnarray}

For a uniform medium one obtains a solution for $\nu(t)$ similar
to eq.(\ref{vacuum}), but with $U$ replaced by the unitary matrix
$\tilde U$ which diagonalizes $\tilde A$, that is, $\tilde A =
\tilde U \hat A \tilde U^\dagger$, and $E-m^2_i/2E$ by $\tilde E -
\lambda_i/2E$ with $\lambda_i/2E$ being the eigenvalues of $\tilde
A$. The eigenvalues are given by
\begin{eqnarray}
&&\lambda_1 = 2s\cos (\theta/3),\;\; \lambda_2 = 2s \cos(\theta/3
+
2\pi/3),\;\; \lambda_3 = 2s\cos(\theta/3 -2\pi/3),
\;\;\theta = \arccos (t^3/s^3),\nonumber\\
&&2t^3 = \mbox{Det}(2E\tilde A),\;\;3 s^2 = |a_{12}|^2 +
|a_{13}|^2 +|a_{23}|^2 +  a^2_{11}+ a^2_{22} +  a_{11}  a_{22}.
\end{eqnarray}
In this case all phases are dynamic again. No Berry phases are generated.

In a non-uniform medium, a Berry phase may be generated. In this case the
matrix $\tilde U$ is now $x$ dependent through its dependence on
$\rho(x)$. Writing the wave function $\tilde \psi(x)$ in the
diagonal basis of $\tilde A$, we have
\begin{eqnarray}
i{d\over d x} \psi^m = (\hat A - i \tilde U^\dagger {d\over dx}
\tilde U) \psi^m,
\end{eqnarray}
where $\psi^m = \tilde U^\dagger \tilde \psi$.

Taking the second term on the right in the above equation as
perturbation, in the adiabatic approximation, one obtains
\begin{eqnarray}
\nu(t,x)_i = e^{-iEt}\sum_j \tilde U_{ij}e^{+i \gamma_j}
e^{i\int^x_0 (\tilde E - \lambda_j/2E)dx} \nu_j^m(0),
\end{eqnarray}
where $\gamma_j$ is given by

\begin{eqnarray}
\gamma_j = i\int_0^L (\tilde U^\dagger{d\over dx} \tilde U)_{jj}
dx = i\int_C (\tilde U^\dagger \vec \bigtriangledown \tilde
U)_{jj} d\rho. \label{phase}
\end{eqnarray}
In the above we have allowed (as will be important later) for the
density parameter to be multi-dimensional. ``$\vec
\bigtriangledown$'' is the gradient taken in density parameter
space $\vec \rho$. The position L is chosen so that $\vec \rho(0)
= \vec \rho(L)$, following Berry's prescription, and $C$ is a
closed curve in parameter space. This is the Berry phase in
neutrino oscillation.

\section{ Conditions for A Non-zero Berry Phase}

Using the above definition of Berry phase in neutrino oscillation,
certain conditions have to be satisfied to generate a non-zero
Berry phase. We now discuss these conditions.

The matrix $\tilde U$ must contain complex phases in order to have
a non-zero $\gamma_j$. Therefore there must be at least three
neutrino mixing to have a Berry phase since with two neutrino
mixing with just W and Z interactions, the matrix $U$ can always
be made real.

A non-zero $(\tilde U^\dagger \bigtriangledown_\rho \tilde
U)_{jj}$ is not a sufficient condition for a non-zero $\gamma_j$.
It is important to realize that the matrix $\tilde U$ is not
uniquely defined. Because the matrix $\hat A$ is diagonal, if
$\tilde P = \mbox{diag}(e^{i\theta_1}, e^{i\theta_2},
e^{i\theta_3})$, then
\begin{equation}
\tilde P \hat A \tilde P^* = \hat A, \quad \mbox{thus}
\quad \tilde A = \tilde U \hat A \tilde U^\dagger =
\tilde U \tilde P \hat A \tilde P^* \tilde U^\dagger.
\end{equation}
Thus both $\tilde U$ and $\tilde U' = \tilde U \tilde P$
diagonalize the matrix $\tilde A$, but as
\begin{eqnarray}
(\tilde U^{'\dagger} \vec \bigtriangledown \tilde U{'})_{jj} =
(\tilde U^\dagger \vec \bigtriangledown  \tilde U)_{jj} + i\vec
\bigtriangledown \theta_j\;, \label{arbphase}
\end{eqnarray}
after integrating over the closed loop $C$, the term proportional
to $\vec \bigtriangledown \theta$ vanishes. Thus the Berry phase
is independent of the choice of the matrix $\tilde U$. One may
regard the transformation $\tilde U \to \tilde U P$ as a gauge
transformation, and the Berry phase is a gauge independent
observable.

The phase $\theta_j$ can be viewed as a pure gauge transformation
of Berry phase. If one can find a gauge in which the Berry phase
is zero before integrating over $C$, the phase is not physical and
can be gauged away.

We illustrate this phenomenon by considering the case of
 three generations.  One way to   obtain the matrix
$\tilde U$ is by requiring
\begin{eqnarray}
(2E \tilde A -\lambda_i) \left (
\begin{array}{l} \tilde U_{1i}\\ \tilde U_{2i}\\ \tilde U_{3i}
\end{array}
\right )=0.\label{ev}
\end{eqnarray}

For the $AN$ given in eq.(\ref{matrix}),
using the first two rows in eq.(\ref{ev}),
we can write

\begin{eqnarray}
\left (
\begin{array}{l} \tilde U_{1i}\\\tilde U_{2i}\\\tilde U_{3i}
\end{array}
\right ) =  {1\over N_i } \left (
\begin{array}{l} ( a_{22} - \lambda_i) a_{13} - a_{23} a_{12}\\
( a_{11}-\lambda_i) a_{23} - a_{13} a^*_{12}\\
a_{12}a_{12}^* - ( a_{22}-\lambda_i) ( a_{11} - \lambda_i)
\end{array}
\right ).\label{c1}
\end{eqnarray}
Here $N_i$ is the normalization constant. One obtains

\begin{eqnarray}
&&(\tilde U {d\over dx} \tilde U^\dagger)_{ii} = -i{1\over N_i^2}
\mbox{Im}(a_{13}a_{12}^*a_{23}^*) {d\over dx} (  a_{22} -
a_{11})\nonumber\\
&& = i \Delta m^2_{21}\Delta m^2_{31}\Delta m^2_{32}
\mbox{Im}(U_{21}U_{23}^*U_{33}U^*_{31})  {1\over N^2_i}{d\over dx}
( a_{22} -  a_{11}).\label{as1}
\end{eqnarray}
Since $d(a_{22}-a_{11})/dx = -(8\sqrt{2}/3)EG_F d\rho/dx\neq 0$,
it seems that a non-zero Berry phase has been generated.


On the other hand, one can also obtain $\tilde U_{ij}$ by using the last two rows of
eq. (\ref{ev}) to obtain
\begin{eqnarray}
\left (
\begin{array}{l} \tilde U'_{1i}\\\tilde U'_{2i}\\\tilde U'_{3i}
\end{array}
\right ) =  {1\over \tilde N_i } \left (
\begin{array}{l}
a_{23}a_{23}^* - ( a_{22}-\lambda_i) ( a_{33} -
\lambda_i)\\
( a_{33} - \lambda_i) a_{12}^* - a_{23} a_{13}^*\\
( a_{22}-\lambda_i) a_{13}^* - a_{12}^* a^*_{23}
\end{array}
\right ).\label{c2}
\end{eqnarray}

One then obtains

\begin{eqnarray}
&&(\tilde U' {d\over dx} \tilde U^{'\dagger})_{ii} = -i{1\over
N_i^2} \mbox{Im}(a_{13}a_{12}^*a_{23}^*) {d\over dx} (  a_{33} -
 a_{22})\nonumber\\
&& = i\Delta m^2_{21}\Delta m^2_{31}\Delta m^2_{32}
\mbox{Im}(U_{21}U_{23}^*U_{33}U^*_{31})  {1\over \tilde
N^2_i}{d\over dx} ( a_{33} -  a_{22}).\label{as}
\end{eqnarray}
For the $AN$ given before $d(a_{33}- a_{22})/dx = 0$,
this gives a zero Berry phase.

However, when we integrate the non-zero result in eq. (\ref{as1})
from $x=0$ to $x=L$ the result is proportional to $\rho(L) -
\rho(0)$ which vanishes. No Berry phase can be generated with one
varying density.

In order to generate a non-zero Berry phase, one must go to a
multi-dimensional parameter space.  This can be achieved if
$a_{ii}$ depends on more than one density $\rho_i$. This may
happen if neutrinos interact with quarks through beyond the SM
interactions, or if there are sterile neutrinos mixed with active
neutrinos,  and therefore the oscillation in medium depends on the
neutron density also. In some astronomical and cosmological
circumstances, there may be significant electron neutrino
densities \cite{TM}, as well as a background electron densities.
Although the neutrino-neutrino interactions are generated by $Z$
exchange, when electron neutrinos interact with electron neutrinos
both direct and exchange diagrams contribute, while when neutrinos
of other flavors interact with the electron neutrino background,
only the direct diagram is possible. Thus the neutrino-background
neutrino interaction is not proportional to a unit matrix in $N$.
In the early universe, significant muon and tauon densities may
occur. For all of these cases that the interaction matrix
depends on more than one density.  The Berry phase developed can
then be written as,
\begin{eqnarray}
\gamma_j = i\int_{\vec \rho(0)}^{\vec \rho(L)}  (\tilde U^\dagger \vec
\bigtriangledown \tilde U)_{jj}\cdot d\vec\rho.
\label{phase1}
\end{eqnarray}
After a cyclic motion, one obtains
\begin{eqnarray}
&& \gamma_j = i\oint (\tilde U^\dagger
\vec \bigtriangledown \tilde U)_{jj}\cdot d\vec \rho
= i\int_{S} \vec \bigtriangledown \times (\tilde U^\dagger
\vec \bigtriangledown  \tilde U)_{jj} \cdot d\vec S,
\nonumber\\
&&= i\int_{S} (\vec \bigtriangledown \tilde U^\dagger\times
\vec \bigtriangledown  \tilde U)_{jj} \cdot d\vec S,
\end{eqnarray}
where $\vec S$ indicates the area enclosed by the path $\vec \rho$
after a cyclic motion. Note that any pure gauge terms are
eliminated since a curl is taken, and $\vec \bigtriangledown
\times \vec \bigtriangledown \theta_j =0$.   Alternatively, we may
say that a non-zero value of $\gamma_j$ indicates that the
function $(\tilde U^\dagger \vec \bigtriangledown \tilde U)_{jj}$
is not a perfect derivative. We find the Berry phase after a
cyclic motion to be given by, for both of our examples in eqs.
(\ref{c1}) and (\ref{c2}),
\begin{eqnarray}
\gamma_i = {2\over 3}\Delta m^2_{21}\Delta m^2_{31} \Delta
m^2_{32} \mbox{Im}(U_{21}U^*_{23}U^*_{31}U_{33}) \int_{S} {\lambda_i
(\vec \bigtriangledown  a_{33}\times \vec \bigtriangledown
a_{22})\cdot d\vec S \over (\lambda_i^2 -s^2 )^3}.
\end{eqnarray}
The above result agrees with that obtained in Ref.\cite{naumov}.

Assuming the interaction matrix $AN$ depends on two varying
independent densities, $\rho$ and $\rho'$, with the entries
$\sqrt{2} G_F ( \alpha_{ii} \rho + \alpha'_{ii} \rho')$
(traceless), we have

\begin{eqnarray}
(\vec \bigtriangledown  a_{33}\times \vec \bigtriangledown
a_{22})\cdot d\vec S &=& (2\sqrt{2} E G_F)^2
(\alpha_{11}\alpha'_{22}- \alpha_{22}\alpha'_{11}) d S.
\end{eqnarray}

It is clear, from the above discussions, that in order to have
non-zero values for the Berry phase $ \gamma_i$ in neutrino
oscillation, there must be at least three generations of neutrinos
with non-vanishing CP violating phases, and the neutrino must
propagate through a background medium with which it interacts
through at least two independent densities.

\section{ Majorana Neutrino Oscillation}

We have seen in the previous discussions that CP violating phase
in the mixing matrix plays an important role in generating a
non-zero Berry phase. In the case of Dirac neutrino this requires
that there are at least three generations of neutrinos.
In the case of Majorana neutrino, there are CP
violating phases even with two generations.
One might naively expect a non-zero Berry phase with two generations.
We now clarify whether CP violating
Majorana phases can play a role in generating
a non-zero Berry phase.
The relevant Lagrangian for
Majorana neutrinos is given by
\begin{eqnarray}
L &=& \bar \nu_L i \gamma^\mu \partial_\mu \nu_L -{1\over 2} \bar
\nu_L^c M \nu_L - {1\over 2} \bar \nu_L M^\dagger \nu^c\nonumber\\
 &-&{g\over 2
\cos\theta_W } \bar f \gamma_\mu(g^f_V - g^f_A\gamma_5) f Z^\mu
 - {g\over \sqrt{2}} \bar l_L
\gamma^\mu \nu_L W_\mu^- + h.c.,\label{mm}
\end{eqnarray}
where $\nu_L^c = C \bar \nu_L^T$ and $C = i\gamma^2\gamma^0$.

The mass matrix $M$ is symmetric in this case, and can be
diagonalized in the following way $
M = \sigma^* U^* P^* \hat M P^* U^\dagger \sigma^*$,
where $\sigma=\mbox{diag}(e^{i\beta_1}, e^{i\beta_2}, e^{i\beta_3})$ and
$P = \mbox{diag}(1,e^{i\alpha_2},e^{i\alpha_3})$ are diagonal phase
matrices. The phases in $\sigma$ can be absorbed into redefinition
of charged lepton fields. In the neutrino mass eigenstate basis,
the charged current interaction term can be written as $
-(g/\sqrt{2}) \bar l_L \gamma^\mu UP \nu_L W^-_\mu$.

When neutrinos pass through a medium, an interacting term
$-\sqrt{2} G_F j_\mu \bar \nu_L \gamma^\mu N \nu_L$ needs to be
added to the Lagrangian. One obtains the equations of motion as,
\begin{eqnarray}
&&i\gamma^\mu \partial_\mu \nu_L - M^\dagger \nu^c_L - \sqrt{2}
G_Fj_\mu \gamma^\mu N \nu_L =0, \nonumber\\
&& i\gamma^\mu \partial_\mu \nu_L^c - M\nu_L  + \sqrt{2}G_F j_\mu
\gamma^\mu N \nu_L^c =0.
\end{eqnarray}

 Expressing the above equations
in the form involving just $\nu_L$, we have
\begin{eqnarray}
&&[-\partial^2  -M^\dagger M - \sqrt{2} G_F i \gamma^\mu
\partial_\mu (j_\nu \gamma^\nu N)\nonumber\\
&& + \sqrt{2} G_F j_\nu \gamma^\nu M^\dagger N (M^\dagger)^{-1}
(i\gamma^\mu \partial_\mu - \sqrt{2}G_F j_\mu \gamma^\mu N)]\nu_L
= 0. \label{majorana}
\end{eqnarray}
In the above we have assumed that none of the neutrinos has zero
mass. We have in mind to see if there are just two generations of
neutrinos a non-zero Berry phase can be generated. With two
generations, if one of the neutrinos has zero mass, the Majorana
phases in $P$ can be completely removed and therefore no Berry
phase can be developed. We need to discuss the case where $M^{-1}$
exists.

The first three terms in eq. (\ref{majorana}) are the same as the
equation of motion for Dirac neutrinos. Since $M^\dagger M = UP
\hat M^* PU^T*U^*P^* \hat M P^* U^\dagger = U\hat M^2 U^\dagger$,
the Majorana phases do not appear in the first three terms. The
Majorana phases may appear in the additional two terms. However,
we note that the term $(i\gamma^u\partial_\mu - \sqrt{2} G_F
\gamma^\mu j_\mu N)\nu_L$ is of order $M^\dagger \nu_L^C$.
Compared with the third term there is a suppression factor $M/E$.
For practical applications, $M/E$ is much smaller than one and can
be safely neglected. With this approximation, one therefore
concludes that no effect of Majorana phases will show up in
neutrino oscillations. One obtains the same equation of motion for
Majorana neutrinos as for Dirac neutrinos with the same
approximation: CP violating Majorana phases do not play a role in
neutrino oscillation\cite{bilenky,langacker}. We do not agree with
the equation of motion for Majorana neutrinos obtained in
Ref.\cite{wrong1}.

\section{ Active-Sterile Neutrino Oscillation}\label{sterile}

The above discussions clearly show that in order to have a
non-zero Berry phase, there must be at least three generations of
neutrinos no matter whether they are Dirac or Majorana neutrinos.
Can the situation be changed with further modifications? In the
following we consider two examples where a non-zero Berry phase
can be developed with just two active neutrinos.

Our first example involves active and sterile neutrino
oscillations. Light sterile neutrinos $\nu^i_R$ may be needed if
the LSND result\cite{lsnd} for neutrino oscillation is confirmed.
The Lagrangian describing light left-handed active and light
sterile neutrinos is similar to eq. (\ref{mm}), but with $\nu_L$
replaced by $(\nu_L , \nu_R^c)^T$ and the mass matrix $M$ replaced
by

\begin{eqnarray}
M= \left (
\begin{array}{ll}
M_{LL}&M_D^T\\ M_D&M_{RR}
\end{array}
\right )\,
\end{eqnarray}
where the different terms are defined by terms in the Lagrangian:
$-(1/2)\bar \nu_L^c M_{LL} \nu_L$, $-\bar \nu_R M_D \nu_L$ and
$-(1/2)\nu_R M_R \nu_R^c$.

The matrices $N$  and $N'$ are still diagonal, with $N =
diag(1,0,0,...,0)$, and $N' = diag(I_{n_a},O_{n_s})$. Here $n_a$
and $n_s$ are the numbers of the active and sterile neutrinos. The
$n\times n$ matrices $I_{n}$ and $O_n$ are a unit matrix and a
matrix with all elements equal to zero, respectively. The matrix
$N$ plays the same role as discussed earlier. The matrix $N'$
which was ignored can no longer be ignored because it is not a
unit matrix and will affect mixing in matter\cite{langacker}. To
have some specific idea on how $N'$ affects oscillation, let us
consider a simple case with two active and one sterile neutrino
oscillation. This is effectively a three generation oscillation
with the $N'$ term included.

One can obtain the equations of motion for the present case by
replacing several quantities in relevant equations. These are

\begin{enumerate}
\item replacing $AN$ in eq. (\ref{medium}) by $AN + B N'$, where
$B =\sqrt{2}G_F \rho'$ with $\rho' = \sum_f \rho_f(T^f_3 - 2
\sin^2\theta_W Q_f) = (-1/2+2 \sin^2\theta_W) \rho +
(1/2-2\sin^2\theta_W)\rho_p + (-1/2)\rho_n$ with $\rho$, $\rho_p$
and $\rho_n$ being the background electron, proton and neutron
number densities, respectively;

\item replacing $A/3$ in eq. (\ref{matrix}) by $A/3 + n_a B/3$ in
the expression for $\tilde E$;

\item replacing the term in $2E\tilde A$ proportional to $A$ by
$2\sqrt{2}EG_F(AN_{ij} + BN'_{ij} - (1/3)(A+n_a B)\delta_{ij}$.

\end{enumerate}

In particular,

\begin{eqnarray}
a_{ij} &=& {1\over 3}(\Delta m^2_{21} - \Delta
m^2_{32})\delta_{ij} - \Delta m^2_{21} U_{i1}U^*_{j1} +\Delta
m^2_{32} U_{i3}U^*_{j3}\nonumber\\
 &+&{1\over 3} 2E \sqrt{2} G_F
(2\delta_{i1}\delta_{j1}(\rho+{1\over 2}\rho')
-\delta_{i2}\delta_{j2}(\rho-
\rho')-\delta_{i3}\delta_{j3}(1+2\rho')).
\end{eqnarray}

An interesting feature in the present case is that in the diagonal
entries $a_{ii}$ of the Hamiltonian, more than one densities
naturally appear which is a necessary condition for Berry phase.
Assuming two active and one sterile neutrino mixing with the
neutron density to be independent from the electron density in a
neutral medium, like the sun and the earth, we obtain
\begin{eqnarray}
(\vec \bigtriangledown  a_{33}\times \vec \bigtriangledown
a_{22})\cdot d\vec S &=& (2\sqrt{2} E G_F)^2 {1\over 6} d S.
\end{eqnarray}
Here the surface is the one spanned by the densities $(\rho,
\rho_n)$ for the closed loop ``C''.

It is possible to have a non-zero Berry phase in oscillations of
two active and one sterile neutrinos, as long as there are CP
violating phases in the mixing matrix. One may wonder if there is
a Berry phase with one active and two sterile neutrino mixing. The
result is negative. In both cases, two active and one sterile, and
one active and two sterile neutrino mixing cases, the mixing
matrix $\tilde U$ contains CP violating phases. However the later
case has $a_{22} = a_{33}$ which results in a zero Berry phase as
can be seen from eq. (\ref{as}). Obviously with more numbers of
active and sterile neutrinos mixing, the conditions for generating
a non-zero Berry phase can be realized.

\section{New Interactions}

We finally consider a case with new interactions. These
interactions may be CP violating and flavor changing. A possible
form of interaction interesting to us is R-parity violating
interactions\cite{rparity}, ${1\over 2}\lambda_{ijk} L^i_L L^j_L
E^{C,k}_R,\;\;\lambda^\prime_{ijk}L_L^iQ^j_L D^{ck}_R$, where
$L_L$, $E_R$, $Q_L$ and $D^c_R$ are the chiral lepton doublet,
charged lepton singlet, quark doublet and down quark singlet in
the supersymmetric extension of SM. The quark super-multiplets are
also color triplet. i, j, and k are flavor indices.
$\lambda_{ijk}$ is anti-symmetric in exchanging the first two
indices. This interaction will not change the PMNS mixing matrix
and neutrino masses, but will change the neutrino interaction in
matter. For example, exchange of right-handed sleptons and squarks
can generate an interaction Lagrangian given by
\begin{eqnarray}
L_{int} = {\lambda_{1ik}\lambda^*_{1jk}\over 2 m^2_{\tilde e_R^k}}
\bar \nu_L^j \gamma^\mu \nu^i_L \bar e_L \gamma_\mu e +
{\lambda'_{i1k}\lambda'^{*}_{j1k}\over 2 m^2_{\tilde d_R^k}} \bar
\nu_L^j \gamma^\mu \nu^i_L \bar d_L \gamma_\mu d_L.
\label{rparity}
\end{eqnarray}
The first term in the above Lagrangian has the same form as
exchange of W between electron and neutrino. The second term is
different. When neutrinos are passing through matter, this term
will generate a term proportional to the neutron and/or proton
density in the Hamiltonian. The neutron density can be independent
from the electron density. For two generation case, the
interaction matrix $AN$ defined earlier is modified to be

\begin{eqnarray}
AN &=& \left ( \begin{array}{ll} A_{11}&A_{12}\\
A_{12}^*&A_{22}\\
\end{array}
\right ) = \sqrt{2} G_F  \left ( \begin{array}{ll} \rho \alpha_{11} +
\rho' \alpha'_{11}& \rho \alpha_{12} e^{i\delta_{12}} +\rho'
\alpha'_{12} e^{i\delta'_{12}}\\\rho \alpha_{12} e^{-i\delta_{12}} + \rho'
\alpha'_{12} e^{-i\delta'_{12}}&\rho \alpha_{22} + \rho' \alpha'_{22}
\end{array}
\right ),
\end{eqnarray}
where $\alpha_{ij} = |\lambda_{1ik}\lambda^*_{1jk}|/4\sqrt{2}G_F
m^2_{e_R^k}$, and $\delta_{12}$ is the phase of
$\lambda_{1ik}\lambda^*_{1jk}/m^2_{e_R^k}$. Similarly for the
primed quantities. CP is violated if $\sin\delta_{12} \neq 0$ or
$\sin\delta'_{12}\neq 0$. $\rho$ is the electron density in the matter.
$\rho'$ includes proton and neutron densities in matter which come
from the second term in the above Lagrangian. $\rho'$ can be
independent from $\rho$. There are constraints on the allowed size
for the R-parity violating interactions\cite{rparity}.
$\alpha_{ij}$ and $\alpha'_{ij}$ can be as large as a percent.
Since our purpose is to demonstrate the possibility of having a
non-zero Berry phase with just two neutrino generation, the
actual number is not important for us.

The quantities $\tilde E$ and  $\tilde A$ in eq.(\ref{matrix})
for the above case are given by

\begin{eqnarray}
\tilde E &=& E - {1\over 2} (m^2_1 + m^2_2) - {1\over 2} (A_{11} +
A_{22}),\;\;
\tilde A = {1\over 2 E} \left ( \begin{array}{ll}
a_{11}& a_{12}\\
a_{12}^*& -a_{11}
\end{array}
\right ),\nonumber\\
a_{11} &=& -{1\over 2} \Delta m^2_{21} \cos(2\theta) - {1\over 2}
2
E(A_{22}-A_{11}),\nonumber\\
a_{12} &=& {1\over 2} \Delta m^2_{21} \sin (2\theta) + 2E A_{12}.
\end{eqnarray}
where $\theta$ is the vacuum mixing angle.
The eigenvalues of the
matrix $2E \tilde A$ are $\lambda_1 =
\sqrt{a^2_{11}+|a_{12}|^2}$ and $\lambda_2 = - \sqrt{a^2_{11}+ |a_{12}|^2}$.

We find
\begin{eqnarray}
\gamma_j &=& i \oint (\tilde U^\dagger \vec \bigtriangledown
\tilde
U)_{jj}\cdot d \vec \rho = i\oint {a^*_{12}\vec \bigtriangledown
a_{12} - a_{12}\vec \bigtriangledown a^*_{12}
\over 4 \lambda_j (\lambda_j - a_{11})}\cdot d \vec \rho\nonumber\\
&=& - i\int_{S}{1\over 4 \lambda^3_j} [ a_{11}\vec
\bigtriangledown a^*_{12}\times \vec \bigtriangledown a_{12} +
\vec \bigtriangledown a_{11} \times (a_{12}\vec \bigtriangledown
a^*_{12} - a^*_{12}\vec \bigtriangledown a_{12})]\cdot d\vec
S \nonumber\\
&+&{i\over 4} \int_S {|a_{12}|^2\over \lambda_j(\lambda_j - a_{11})}
\vec \bigtriangledown \times {a^*_{12}\vec \bigtriangledown a_{12}
-a_{12}\vec \bigtriangledown a^*_{12}\over |a_{12}|^2}\cdot d\vec S .\nonumber\\
\end{eqnarray}

To demonstrate that indeed the above can produce a non-trivial
geometric phase, let us consider a simple case where $a_{11} = a$
and $a_{12} = b e^{i\omega x}$ with $a$ and $b$ constants.
Integrating over one period, $x = 2\pi/\omega$, the geometric
phase is given by
\begin{eqnarray}
\gamma_1 = -\pi (1 + {a\over \sqrt{a^2+b^2}})\;,\;\;\;\;
\gamma_2 = -\pi (1 - {a\over \sqrt{a^2+b^2}})\;\;.
\end{eqnarray}
Note that in the above example CP is violated because $a_{12}$ is complex.

We see that a non-zero geometric Berry phase can be developed in
two neutrino oscillation if there are at least two independent
varying matter densities, and also a non-zero CP violating phase
difference in the off diagonal elements of the interaction matrix
$AN$. The case can be easily generalized to the case with three
neutrinos.

\section{ Conclusions}

We have studied Berry phase in neutrino oscillations for both
Dirac and Majorana neutrinos. In order to have  a non-zero Berry
phase there must exist at least three generations of neutrinos
with CP violation in the mixing matrix and the oscillation must
occur in a background with more than one varying  densities if the
interaction with matter is due to the standard  model W and Z
exchange. CP violating Majorana phases do not play a role in
generating a Berry phase.

 If neutrino oscillations involve only
active neutrinos, the interaction of Z boson exchange does not
affect neutrino oscillations in matter. If there is active and
sterile neutrino mixing, the situation changes and Z boson
exchange does affect neutrino oscillations in matter. This
scenario provides a natural setting to realize the condition of
more than one density in matter since the Z boson exchange is
sensitive to electron and also neutron densities in matter.

With new CP violating and flavor changing interactions, it is
possible to have non-zero Berry phases even with just two
generations, but the interactions must still depend on two
independent densities in the background medium.

\section*{Acknowledgements}
This work was supported in part by the NSC, NNSFC, and ARC. XGH
and BMcK are, respectively, grateful to the School of Physics,
University of Melbourne and the Departments of Physics at Nankai
University and National Taiwan University for their hospitality
where parts of the work reported here were done.

\end{document}